# Role of the Non-Collinear Polarizer Layer in Spin Transfer Torque Switching Processes

Chun-Yeol You[1] and Myung-Hwa Jung[2,a)]

[1]Department of Physics, Inha University, Incheon 402-751, Korea

[2]Department of Physics, Sogang University, Seoul 121-742, Korea

We have recently reported that the spin transfer torque switching current density is very sensitive to not only the junction sizes but also the exchange stiffness constants of the free layer according to the micromagnetic simulations. The results are very complicate and far from the macro-spin model because of the non-coherent spin switching processes. The dependence of the switching current density on the junction sizes and the exchange stiffness constants becomes systematic when we employ the non-collinear polarizer layer. It is found that the non-collinear polarizer layer enhances the coherency of the spin dynamics by breaking symmetric spin configurations.

PACS: 75.76.+j, 72.25.-b, 85.75.Dd, 75.78.Cd

[a)]Author to whom correspondence should be addressed. E-mail : mhjung@sogang.ac.kr

## I.    Introduction

The switching current density in the spin transfer torque magnetic random access memory (STT-MRAM) is one of the most important key issues in the development of the next generation non-volatile memory devices. A high switching current density causes the reliability problem of magnetic tunneling junction (MTJ) and the size limit of device. Therefore, there have been many approaches to reduce the switching current density through the optimization of layer structures and magnetic properties of the switching layer.[1,2,3,4,5,6,7] Recently, we have reported that the switching current density is very sensitive to the exchange stiffness constants[8] and the lateral junction dimensions[9] by micromagnetic simulations. We have also found that the switching current dependences are unpredictably complicated,[8,9] which cannot be explained with simple model. Since the complicated spin configurations during the switching processes are main reasons of unexpected switching current density dependences, the detail spin configuration can be determined by the characteristic length scale of $\sqrt{A_{ex}/K_{eff}}$, where $A_{ex}$ is the exchange stiffness constant and $K_{eff}$ is an effective anisotropy energy including demagnetization energy. When the lateral dimensions of the junction are comparable with that characteristic length, the detail spin configuration becomes very complex. Therefore, such unexpected switching current density dispersion with small variation of junction sizes and fabrication conditions may act as a hurdle to commercialize the STT-MRAM, and it is clear that more coherent spin rotation is desirable for the device operations.

By virtue of ultrafast time resolution and nanometer scale spatial resolution of the X-ray microscopy, it has been revealed that the complicate non-uniform modes are created during the switching procedures for the $110 \times 180$, $110 \times 150$, and $85 \times 135$ nm$^2$ junctions.[10,11] Even though there are no vortex formations in the smaller junctions unlike $110 \times 180$ nm$^2$ junction, the switching process are always accompanied with complex non-uniform and non-coherent spin configurations. It has been known that the "non-collinear polarizer layer", the magnetization directions between polarizer layer and switching layer are non-collinear, leads better coherent spin rotations.[12,13] It has been shown that the non-collinear polarizer layer leads more coherent spin rotations by experiments and micromagnetic simulations for the $60 \times 130$ nm$^2$ junctions.[13] We also demonstrated that the non-collinear polarizer layer structure can reduce the switching current density significantly (40 %) even for the smaller junction size ($60 \times 40$ nm$^2$).[14] Based on our previous work about the non-collinear polarizer layer, in this work, we would like discuss more details of the role of the non-collinear polarizer layer. We found that the non-collinear polarizer layer plays important roles not only in the reduction of the switching current density, but also in weak and predictable dependences of the exchange stiffness constants and the junction sizes. Therefore, the non-collinear polarizer layer structure has many advantages in the devices designs.

## II. Macro-Spin Model

Figure 1 shows the MTJ structure with the non-collinear polarizer. The magnetization of the switching layer $F_1$ is aligned along the $x$ axis, and the magnetization of the polarizer layer in the SyA structure ($F_2$/NM/$F_3$) is aligned with an angle $\theta_p$ from the $x$ axis. Here, I, NM, and AFM are the insulating, non-magnetic, and antiferromagnetic layers, respectively. The direction of positive current and the axis of coordinates are also depicted in Fig. 1. A, B, C, and D indicate local points of the switching layer, which will be used later.

Let us derive the analytic expression of the switching current density with the non-collinear polarizer layer based on the macro-spin model. The Landau-Lifshitz-Gilbert (LLG) equation with STT term is

$$\frac{d\vec{m}_s}{dt} = -\gamma\,\vec{m}_s\times\vec{H}_{eff} + \alpha\vec{m}_s\times\frac{d\vec{m}_s}{dt} - \gamma\,a_J\vec{m}_s\times(\vec{m}_s\times\vec{m}_p) - \gamma\,b_J(\vec{m}_s\times\vec{m}_p)\,. \qquad (1)$$

Here, $\gamma$ is the gyromagnetic ratio, $\alpha$ is the Gilbert damping constant, and $a_J = a_1 J$ with $a_1 = \eta_p\dfrac{\hbar}{2e\mu_0 M_s d_s}$ and $J$ the current density (opposite to the electron flow). $\vec{m}_{s,p}$ and $\vec{m}_{s,p}$ are the unit vectors of the magnetization of switching and polarizer layers, and $\vec{H}_{eff}$ is the effective field including external field. $\eta_p$, $d_s$, $M_s$, $\mu_0$, and $\hbar$ are the spin polarization of the polarizer layer, the thickness of the switching layer, the saturation magnetization, the permeability of the vacuum, and the reduced Plank's constant, respectively. More details can be found in our previous works.[14,15,16] For simplicity, we ignore the field-like STT term by setting $b_J = 0$ because the overall $J_c$ dependence on the polarizer angle is not changed with the field-like term, as we have already shown in our previous result.[14]

We repeat the standard method of spin wave excitation to find the switching current density by using instability conditions[17,18,19] with the non-collinear polarizer layer. We assume that the initial state is $\vec{m}_s(t=0)=\hat{x}$ and $\vec{m}_p(0)=\cos\theta_p\hat{x}+\sin\theta_p\hat{y}$, where $\theta_p$ is the magnetization angle of the polarizer layer from the $x$ axis as shown in Fig. 1. Let us define the excited non-zero $x$- and $y$-components of magnetization by STT, that is $\Delta\vec{m}_{x,y}(t)$. We put $\Delta\vec{m}_{x,y}(t)$ contributions to Eq. (1), and linearize them up to the first order of $\Delta \overline{m}_{x,y}(t)$, because they are supposed to be small. After linearizing, we obtain two coupled differential equations of $\Delta \overline{m}_{x,y}(t)$. Let us put $\Delta m_{x,y}(t) = \Delta \overline{m}_{x,y} e^{kt}$ with the simple harmonic oscillation model, here $k$ is a complex number. With the same procedures with previous report,[6] we find following matrix equation;

$$\begin{pmatrix} k + \gamma a_J \cos \theta_p & k\alpha + \gamma H_{eff} + \gamma \left( N_y - N_x \right) M_s \\ k\alpha + \gamma H_{eff} + \gamma \left( N_z - N_x \right) M_s & k + \gamma a_J \cos \theta_p \end{pmatrix} \begin{pmatrix} \Delta \overline{m}_x \\ \Delta \overline{m}_y \end{pmatrix} = 0 , \qquad (2)$$

where $N_x$, $N_y$, and $N_z$ are demagnetization factors of each direction. The determinant of matrix must be zero in order to have non-trivial solutions. The characteristic equation of the matrix is a quadratic equation of $k$, and the instability condition is

$$\text{Re}[k] = -\frac{\gamma}{1+\alpha^2} \Big[ \alpha \left( H_{eff} + \left( N_y + N_z - 2N_x \right) M_s/2 \right) + a_J \cos \theta_p \Big] > 0 . \qquad (3)$$

Then, we find $J_c \left( \theta_p \right)$ as follows;

$$a_J \cos \theta_p = a_1 J \cos \theta_p < -\alpha \left( H_{eff} + \left( N_y + N_z - 2N_x \right) M_s/2 \right), \qquad (4)$$

$$J_c \left( \theta_p \right) < -\frac{\alpha}{a_1 \cos \theta_p} \left( H_{eff} + \left( N_y + N_z - 2N_x \right) M_s/2 \right) = \frac{J_{c0}}{\cos \theta_p} . \qquad (5)$$

Here, $J_{c0}$ is the switching current density with the collinear polarizer layer ($\theta_p = 0$) in a normal STT structure. From these results, we can claim that the switching current density of the non-collinear polarizer is inversely proportional to the cosine. Furthermore, we also numerically solve the LLG equation with the macro-spin model,[6] and the results show the inverse cosine dependence as shown in Fig. 2 with Eq. (5). In Fig. 2, we also depict the normalized switching current densities from previous micromagnetic simulation result.[14]

## III.    Micromagnetic Simulations

The micromagnetic simulations are performed by using the public domain micromagnetic simulator, Object-Oriented MicroMagnetic Framework (OOMMF)[20] with the public STT extension module.[15] A typical STT-MRAM structure[21] in Fig. 1 is fully considered in micromagnetic simulations. The saturation magnetization $M_s$ and thicknesses of the $F_1$, $F_2$, and $F_3$ layers are $1.3 \times 10^6$ A/m and 2 nm, respectively. The thicknesses of the NM and I layers are 1 nm. The cross-section of the nano-pillar is an ellipse of $60 \times 40$ nm$^2$ (or varied from 40 to 120 nm $\times$ 30 nm in Sec. V), with a mesh size of $1 \times 1 \times 1$ nm$^3$. For simplicity, the crystalline anisotropy energy is not considered in this study. $A_{ex}$ was set as $2.0 \times 10^{-11}$ J/m (or varied from 0.5 to $3.0 \times 10^{-11}$ J/m in Sec. V), and the Gilbert damping constants $\alpha$ is 0.02. The exchange bias field of $4 \times 10^5$ A/m is assigned to a $\pi + \theta_p$ direction from the long axis of the ellipse ($+x$ direction) for the $F_3$ layer. Due to the exchange bias, the magnetization direction of the $F_3$ layer prefers the $\pi + \theta_p$ direction. Because we design a strong antiferromagnetic interlayer exchange coupling ($-1.0 \times 10^{-3}$ J/m$^2$) between the $F_2$ and $F_3$ layers to mimic a typical SyF structure, the magnetization direction of the $F_2$ layer is $\theta_p$ as shown in Fig. 1. More details of the micromagnetic simulation results have been reported in Ref. [14].

## IV.    Reduction of the Switching Current Density

We depict the results of the switching current densities of the micromagnetic simulation, Eq.(5), and the numerical simulations of the macro-spin model in Fig. 2.

Here, the numerical simulation of the macro-spin means the switching current density is obtained by numerically solving LLG equations for the macro-spin. In Fig. 2, all switching current densities are normalized with the value of the collinear polarizer layer ($\theta_p = 0$). Therefore, it must be mentioned that values $J_{c0}$ are different for three cases; $J_{c0}$ = $2.7 \times 10^{11}$ A/m$^2$ for the micromagnetic simulation, $J_{c0}$ = $1.84 \times 10^{11}$ A/m$^2$ for the analytic macro-spin model Eq. (5), and $J_{c0}$ = $2.17 \times 10^{11}$ A/m$^2$ for the numerical macro-spin calculation, respectively. The reason of such discrepancy will be discussed later.

Surprisingly, the macro-spin model and micromagnetic simulation results are opposite for small $\theta_p$ as shown in Fig. 2. The Eq. (5) and the numerical macro-spin model calculations imply that the non-collinear polarizer will increase the switching current density. However, the micromagnetic simulation results show noticeable reduction of the switching current density. The main reason is that the macro-spin model is failed for the collinear polarizer layer due to the non-uniform spin configuration during switching processes. For the non-collinear polarizer case, the switching processes occur with the coherent spin rotation which is well described by the macro-spin model. Therefore, the reduction of the switching current density for the non-collinear polarizer is mainly due to the enhanced coherent spin rotation.[12,13]

In order to get deep insight of the switching process, we take snapshots of the spin configuration of the switching layer as shown in Figs. 3 and 4 for collinear ($\theta_p = 0$) and non-collinear ($\theta_p = 10^{\circ}$) polarizer layers, respectively. The snapshots are taken for the switching current density of each case. We have already discussed about the detail spin dynamics in our previous reports for the collinear polarizer case.[15,16] Before the actual switching occurs, strong spin oscillations are found at both edges, and the phases of spin oscillations at both edges are opposite.[9] Such asymmetric spin dynamics of both sides are clearly shown in Fig. 3 (a) and (b). In order to check more details, Fig. 3 (e) shows the time dependent normalized magnetization dynamics for the specific positions (A, B, C, and D), which are defined in Fig. 1. In Fig. 3 (e), the amplitude of oscillation is the strongest at the edge (position A), and it decreases from point A (edge) to D (center). Even though there are strong oscillations at both edges, the effect of STT is almost vanished at the center because of the out-of-phase STT contributions from both sides. Therefore, spins at the center position do not oscillate till 8.0 ns. After certain time ($t \sim$ 9.0 ns in Fig. 3 (c)), spins at the center position start to move by breaking the asymmetric spin dynamics. After breaking the asymmetric spin dynamics, the switching is accomplished. It must be emphasized that the detail spin dynamics of the collinear polarizer layer in the typical STT structure is far from the coherent rotation in the macro-spin model.[16]

Now let us consider the non-collinear polarizer case, as shown in Figs. 4 (a)~(d). We find that the spin oscillations are almost coherent from the beginning (Fig. 4 (a)). The rotation angles of each spin are almost similar, and it is quite different for the collinear polarizer case. The coherent rotation of spins is easily confirmed from Fig. 4 (e). Contrary to the collinear polarizer case in Fig. 3 (e), the spin dynamics of each position from A to D for the non-collinear polarizer case are almost identical and in-phased as shown in Fig. 3 (e).

### V.   Dependence of the Lateral Junction Sizes and the Exchange Stiffness Constants

Recently, we have reported that the switching current densities are very sensitive function of the lateral junction sizes[9] and the exchange stiffness constants, $A_{ex}$.[8] As aforementioned, the switching process occurs always with the non-uniform spin configurations, and the detail spin configurations are governed by the exchange length and the lateral dimension of the junction. Since the macro-spin model cannot treat the lateral junction size and the $A_{ex}$,[8,9] the macro-spin model failed to handle such problem properly. Here, we recall our previous results[8,9] for various $A_{ex}$ values of $80 \times 40$ nm$^2$ junctions, and various junction sizes for the collinear ($\theta_p = 0$) polarizer layer. In addition, we repeat the same simulation for the non-collinear ($\theta_p = 10^o$) polarizer layer in Fig 5 (a) and (b). As shown in Fig. 5 (a), the switching current density dependence on $A_{ex}$ is unpredictable for collinear polarizer case. It must be noted that the $A_{ex}$ variation ($1.0 \sim 3.0 \times 10^{-11}$J/m) is realistic for CoFeB fluctuated with the fabrication conditions.[22] Within these realistic variations, the switching current density varies around 60%. However, when we consider non-collinear polarizer layer, the variation of the switching current density shows almost monotonic decrease, and it is predictable. For the larger $A_{ex}$ values, which is more likely to be macro-spin status, the switching current density approaches to the $J_{c0}$ ($= 1.84 \times 10^{11}$ A/m$^2$) value based on the analytic macro-spin model with increasing $A_{ex}$.

In Fig. 5 (b), we fix $A_{ex} = 2.0 \times 10^{-11}$ J/m and short axis of the ellipse, $b = 30$ nm, and varies the long axis of the ellipse, $a$, from 40 to 120 nm. The collinear polarizer shows overall increase, but the data is strongly scattered so that it is not easy to say general trend. However, when we consider the non-collinear polarizer, the dependence is monotonically increased with small deviation. Such predictable behavior implies that more coherent spin switching is insensitive on the lateral dimension of the junction and $A_{ex}$, because non-uniform spin configuration is not involved in these cases.

## VI.  Conclusion

In conclusion, we investigate the role of the non-collinear polarizer layer in the typical STT structure. Based on our micromagnetic simulations, we can claim that the realistic switching process is far from the macro-spin model in a typical STT structure with the collinear polarizer layer. In contrast to the macro-spin model, the important switching process always involves non-uniform and non-coherent spin rotation for the collinear polarizer. Such non-coherent switching can be suppressed effectively with the non-collinear polarizer layer by breaking the asymmetric spin dynamics, and it leads conspicuous switching current density reduction. Furthermore, unpredictable and very sensitive dependence of the switching current density on the lateral junction sizes and the exchange stiffness constants changes to the weak and monotonic dependences.

### Acknowledgements

This work was supported by the NRF funds (Grant Nos. 2010-0023798 and 2010-0022040, Nuclear R&D Program 2012M2A2A6004261) of the MEST of Korea, and by the IT R&D program of MKE/KEIT (10043398).

**Figure Captions**

Fig. 1 MTJ structure with non-collinear polarizer layer. $F_{1,2,3}$ are the ferromagnetic layers, and I, NM, and AFM are the insulating, non-magnetic, antiferromagnetic layers, respectively. The positive current is defined from the bottom to top electrodes, and *x*-axis is the long axis of the ellipse. The average magnetization angel of the polarizer layer, $\theta_p$, are also depicted. A, B, C, and D indicate the specific positions of the switching layer.

Fig. 2 Normalized switching current density obtained from the micromagnetic simulations, macro-spin models, and Eq. (5) as a function of the polarizer layer magnetization direction, $\theta_p$. Here, $J_{c0} = 2.7 \times 10^{11}$ A/m$^2$ for micromagnetic simulation, $J_{c0} = 1.84 \times 10^{11}$ A/m$^2$ for analytic macro-spin model, and $J_{c0} = 2.17 \times 10^{11}$ A/m$^2$ for numerical macro-spin simulation, respectively.

Fig. 3 Snapshots of spin configurations of the switching layer with the collinear ($\theta_p = 0$) polarizer layer at specific time with the switching current density. (a) $t = 6.44$ ns, (b) $t = 6.51$ ns, (c) $t = 8.99$ ns, and (d) $t = 9.36$ ns. (e) Time dependent normalized magnetization at specific positions (A, B, C, and D), and total magnetization with the collinear polarizer layer.

Fig. 4 Snapshots of spin configurations of the switching layer with the non-collinear ($\theta_p = 10^\circ$) polarizer layer at specific time with the switching current density. (a) $t = 4.13$ ns, (b) $t = 6.00$ ns, (c) $t = 6.13$ ns, and (d) $t = 9.60$ ns. (e) Time dependent normalized

magnetization at specific positions (A, B, C, and D), and total magnetization with the non-collinear polarizer layer.

Fig. 5 Switching current density as a function of (a) the exchange stiffness constant, $A_{ex}$, for the 80 x 40 nm$^2$ junctions, (b) the long axis of the ellipse with 30-nm short axis and $A_{ex} = 2.0 \times 10^{-11}$ J/m. The blue and red circles represent the collinear ($\theta_p = 0^o$) and non-collinear ($\theta_p = 10^o$) polarizer layers, respectively.

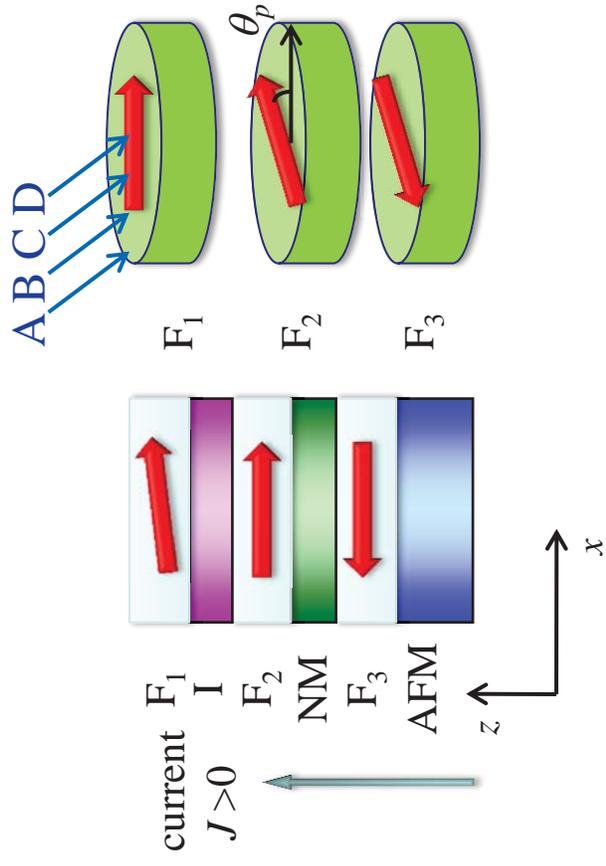

Fig. 1

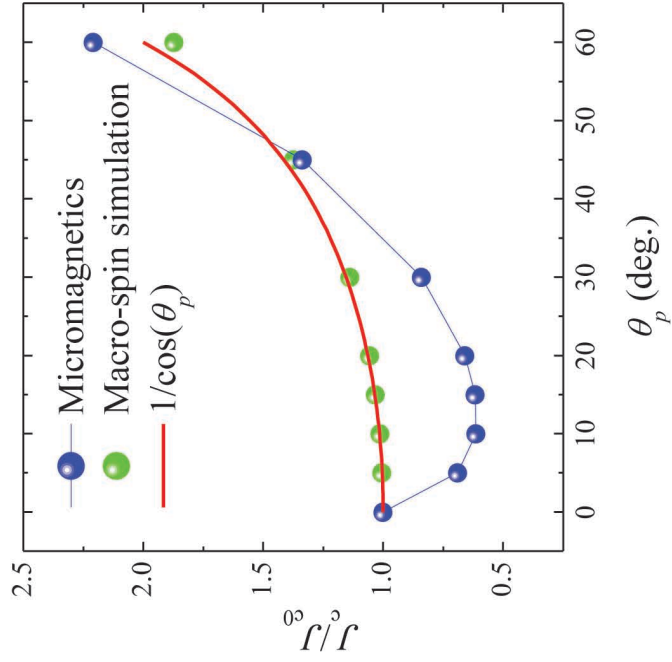

Fig. 2

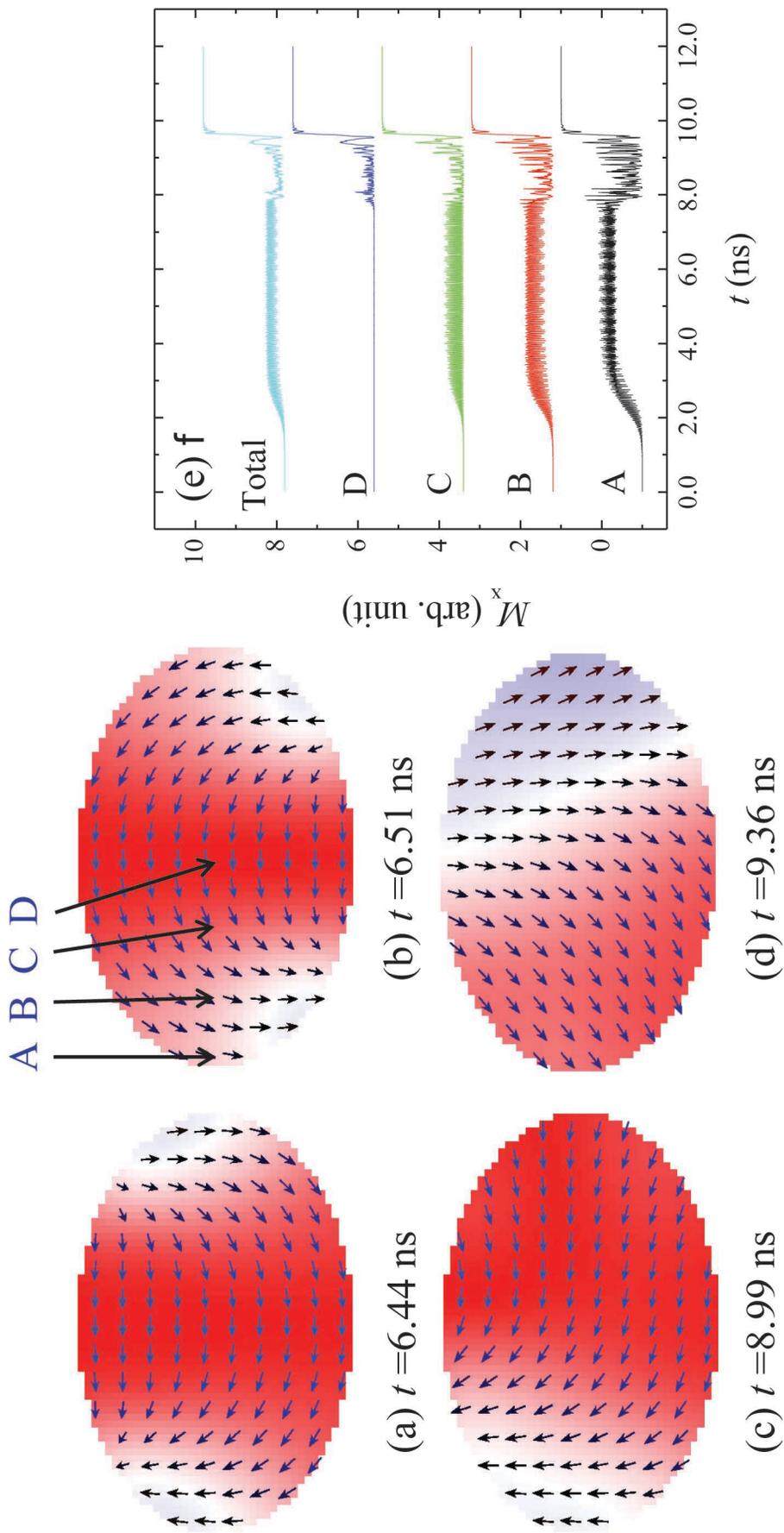

(a) $t = 6.44$ ns

(b) $t = 6.51$ ns

(c) $t = 8.99$ ns

(d) $t = 9.36$ ns

(e) f

Total

D

C

B

A

$M_x$ (arb. unit)

$t$ (ns)

Fig. 3

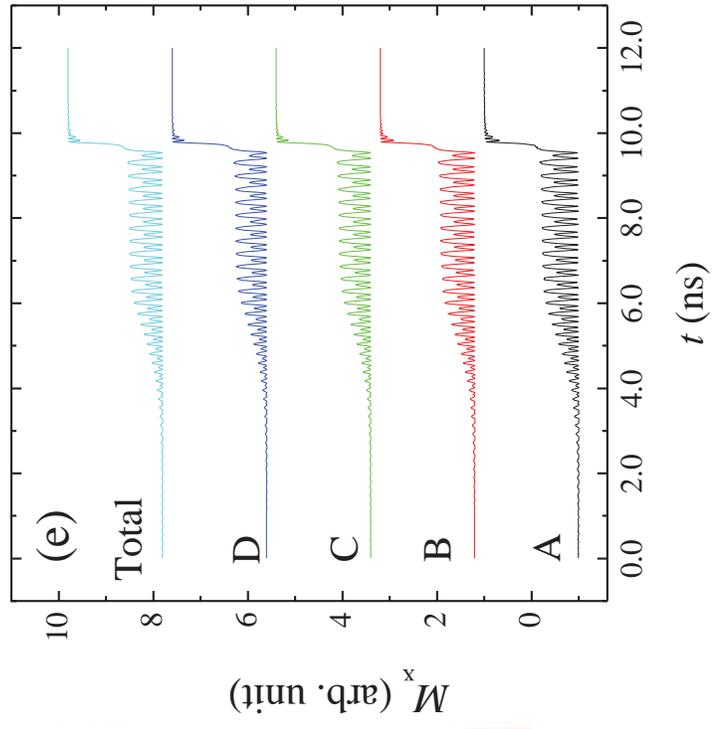

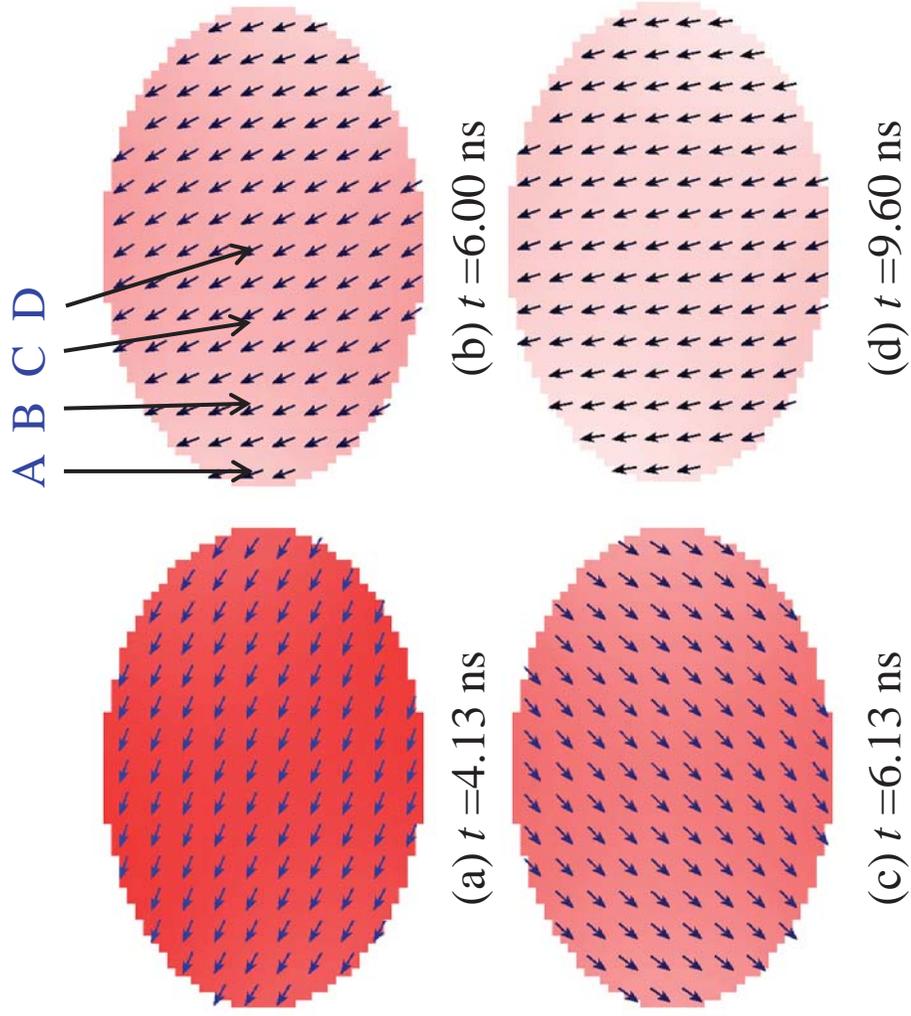

(a) $t = 4.13$ ns

(b) $t = 6.00$ ns

(c) $t = 6.13$ ns

(d) $t = 9.60$ ns

A B C D

(e)

Total

D

C

B

A

$M_x$ (arb. unit)

$t$ (ns)

Fig. 4

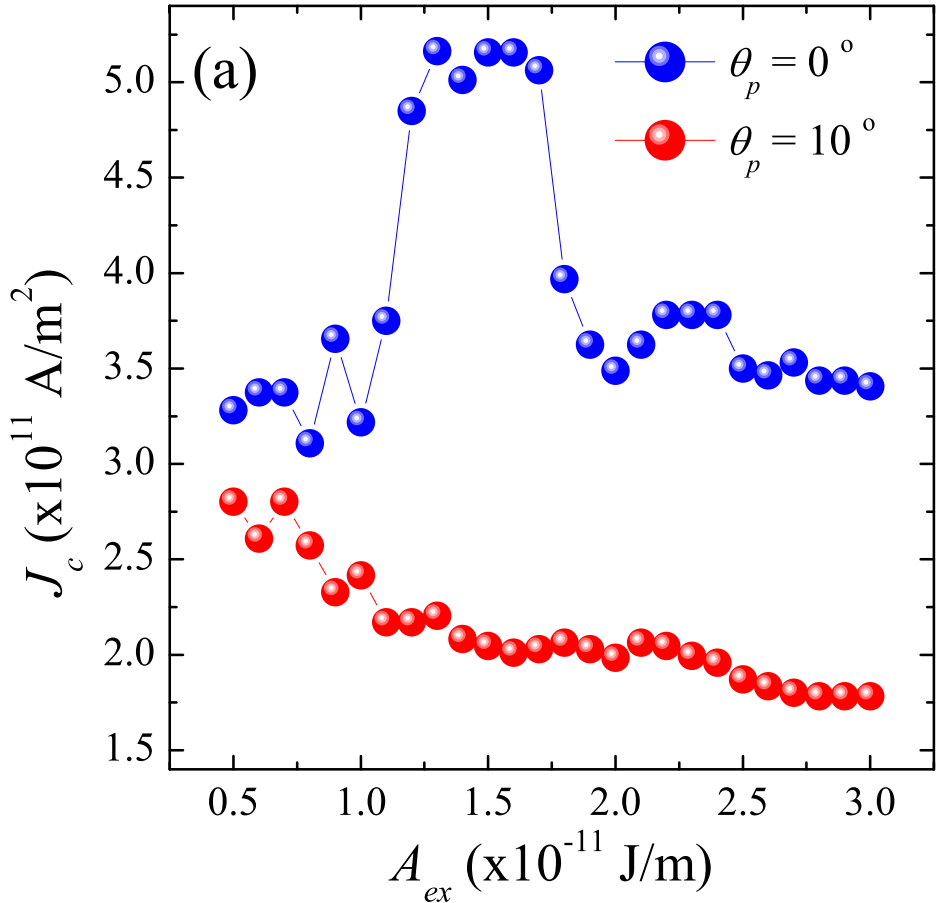

(a)

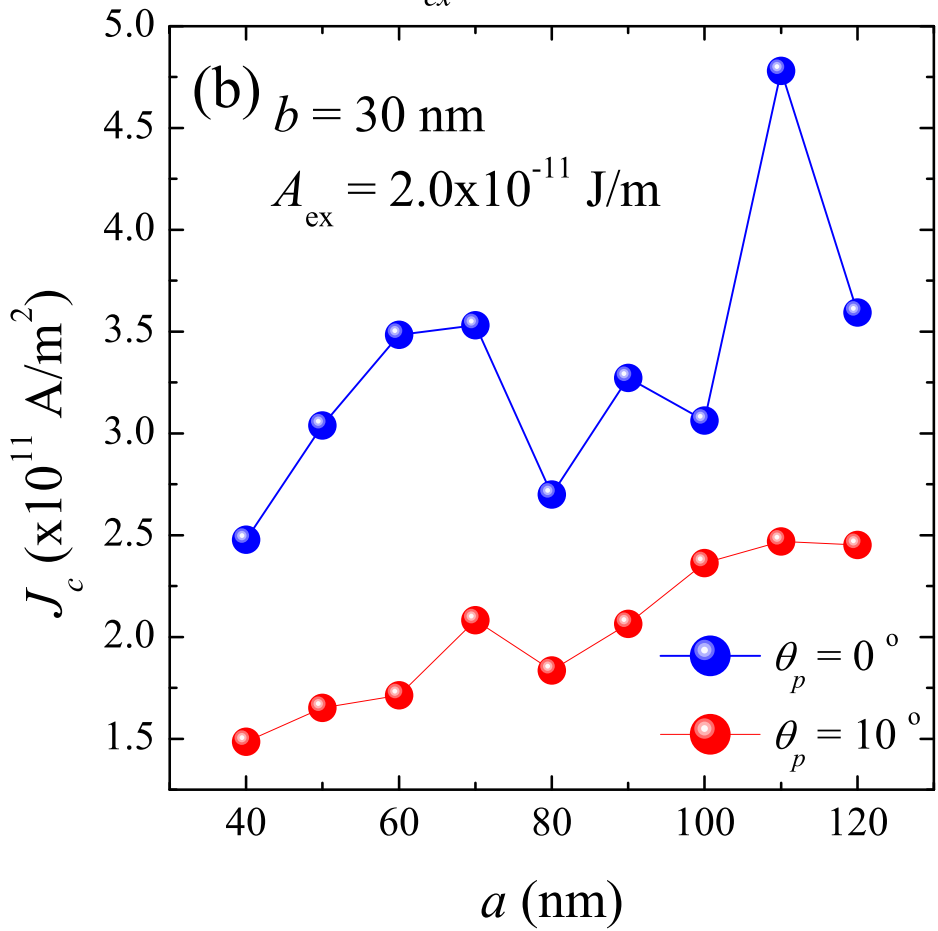

(b)